\documentclass[twocolumn,epjc3]{svjour3}

\usepackage{amsmath}
\usepackage{euscript,amssymb,epsfig}
\usepackage{graphicx}
\usepackage{amsfonts,latexsym}

\RequirePackage{graphicx}

\newcommand{\be}[1]{\begin{equation}\label{#1}}
\newcommand{\ee}{\end{equation}}
\newcommand{\ba}[1]{\begin{eqnarray}\label{#1}}
\newcommand{\ea}{\end{eqnarray}}
\newcommand{\rf}[1]{(\ref{#1})}
\newcommand{\nn}{\nonumber}

\newcommand{\const}{\mbox{\rm const}\,}

\journalname{Eur. Phys. J. C}

\begin{document}

\title{$f(R)$ gravity: scalar perturbations in the late Universe}

\author{Maxim Eingorn\thanksref{e1,addr1,addr2} \and Jan Nov$\mathrm{\acute{\bf a}}$k\thanksref{e2,addr3,addr4} \and
Alexander Zhuk\thanksref{e3,addr5}}

\thankstext{e1}{e-mail: maxim.eingorn@gmail.com}

\thankstext{e2}{e-mail: jan.janno.novak@gmail.com}

\thankstext{e3}{e-mail: ai.zhuk2@gmail.com}

\institute{Physics Department, North Carolina Central University,\\ Fayetteville st. 1801, Durham, North Carolina 27707, U.S.A.\\ \label{addr1} \and Institute
for Theoretical Physics, University of Cologne,\\ Z\"ulpicher st. 77, Cologne 50937, Germany\\ \label{addr2} \and Department of Theoretical Physics, Faculty of
Mathematics and Physics,\\ Charles University, V Holesovickach 2, 180 00, Prague 8, Czech Republic\\ \label{addr3} \and Institute of Mathematics AS CR, Zitna
st. 25, 115 67 Prague 1, Czech Republic\\ \label{addr4} \and Astronomical Observatory, Odessa National University,\\ Dvoryanskaya st. 2, Odessa 65082,
Ukraine\\ \label{addr5} }

\date{Received: date / Accepted: date}

\maketitle

\begin{abstract}
In this paper we study scalar perturbations of the metric for nonlinear $f(R)$ models. We consider the Universe at the late stage of its evolution and deep
inside the cell of uniformity. We investigate the astrophysical approach in the case of Minkowski spacetime background and two cases in the cosmological
approach, the large scalaron mass approximation and the quasi-static approximation, getting explicit expressions for scalar perturbations for both these cases.
In the most interesting quasi-static approximation, the scalar perturbation functions depend on both the nonlinearity function $f(R)$ and the scale factor $a$.
Hence, we can study the dynamical behavior of the inhomogeneities (e.g., galaxies and dwarf galaxies) including into consideration their gravitational
attraction and the cosmological expansion, and also taking into account the effects of nonlinearity. Our investigation is valid for functions $f(R)$ which have
stable de Sitter points in future with respect to the present time, that is typical for the most popular $f(R)$ models.
\end{abstract}


\section{Introduction}

Modern observational phenomena, such as dark energy and dark matter, are the great challenge  for present cosmology, astrophysics and theoretical physics.
Within the scope of standard models, a satisfactory explanation to these problems has not been offered yet. This forces the search of their solutions beyond
the standard models, for example, by considering modified gravitational theories. One of the possible generalizations consists in consideration of nonlinear
(with respect to the scalar curvature $R$) models $f(R)$. Nonlinear models may arise either due to quantum fluctuations of matter fields including gravity
\cite{BirrDav}, or as a result of compactification of extra spatial dimensions \cite{NOcompact}. Starting from the pioneering paper \cite{Star1}, the nonlinear
theories of gravity $f(R)$ have attracted a great deal of interest because these models can provide a natural mechanism of the early inflation. It was also
realized that these models can explain the late-time acceleration of the Universe. This fact resulted in a new wave of papers devoted to this topic (see, e.g.,
the reviews \cite{review1,review2,review,rew3,rew4,rew5}).

The cosmological perturbation theory is very important for the current cosmological investigations of the large-scale structure. Thus, it would be interesting
to make the corresponding cosmological perturbation analysis in nonlinear $f(R)$ theories of gravity. In the hydrodynamical approach, such investigation was
performed in a number of papers (see, e.g., Sec. 8 in the review \cite{review} and references therein). In particular, matter density perturbations in a class
of viable cosmological $f(R)$ models were studied in \cite{Stfl1,Stfl2}. We consider the Universe at the late stage of its evolution when galaxies and clusters
of galaxies have already formed. At scales much larger than the characteristic distance between these inhomogeneities, the Universe is well described by the
homogeneous and isotropic FRW metric. These scales are approximately 190 Mpc and larger \cite{EZcosm2}. At these distances, the matter fields (e.g., cold dark
matter) are well described by the hydrodynamical approach.  However, at smaller scales the Universe is highly inhomogeneous, and we need to take into account
the inhomogeneities in the form of galaxies, groups and clusters of galaxies. The peculiar velocities of these inhomogeneities are much less than the speed of
light, and we can use the nonrelativistic approximation. This means that in equations for scalar perturbations we first neglect peculiar velocities and solve
these equations with respect to scalar perturbation functions $\Phi$ and $\Psi$. The function $\Phi$ represents the gravitational potential of the
inhomogeneities. Then, we use the explicit expression for $\Phi$ to describe the motion of inhomogeneities. Such mechanical approach is well known in
astrophysics (see, e.g., \cite{Landau}). We generalized it to the case of dynamical cosmological background \cite{EZcosm2,EZcosm1}. In the case of the linear
model (i.e. the conventional $\Lambda$CDM model), we used this procedure to describe the mutual motion of galaxies and dwarf galaxies \cite{EZcosm1,EKZ2}. Due
to the great popularity of the nonlinear $f(R)$ models, it is of interest to apply this scheme to them. However, first of all, we should show that nonlinear
theories are compatible with the mechanical approach. In other words, we have to examine equations for scalar perturbations of the metrics in nonlinear $f(R)$
gravity within the framework of the mechanical approach to show their integrability up to the required accuracy. This is the main aim of our paper.

As a result, we demonstrate that considered in our paper nonlinear theories are compatible with the mechanical approach. We get the expressions for both $\Phi$
and $\Psi$ in different approximations. Moreover, the exact form of the gravitational potential $\Phi$ gives a possibility to take into account both the
effects of nonlinearity of the original model and the dynamics of the cosmological background. The explicit form of this function makes it possible to carry
out analytical and numerical study of mutual motion of galaxies in nonlinear models. Therefore, our formulae can be used to analyze the large-scale structure
dynamics in the late Universe for nonlinear f(R) models. This is the main result of our paper.

The paper is structured as follows. In Sec. 2 we present the main background equations as well as the equations for scalar perturbations for an arbitrary
$f(R)$ model. Here the equations for scalar perturbations are written within the framework of the mechanical approach. In Sec. 3 we solve these equations in
three approximations: the astrophysical approach, the large scalaron mass case and the quasi-static approximation. In all three cases we obtain the expressions
for the scalar perturbation functions $\Phi$ and $\Psi$ up to the required accuracy. The main results are summarized in concluding Sec. 4.

%%%%%%%%%%%%%%%%%%%%%%%%%%%%%%%%%%%%%%%%%%%%%%%%%%%%%%%%%%%%%%%%%%%%%%%%%%%%%%%%%%%%%%%%%%%
%%%%%%%%%%%%%%%%%%%%%%%%%%%%%%%%%%%%%%%%%%%%%%%%%%%%%%%%%%%%%%%%%%%%%%%%%%%%%%%%%%%%%%%%%%%

\section{Basic equations}

\setcounter{equation}{0}

In this section we reproduce some known equations of the nonlinear $f(R)$ gravitational model that we will use hereinafter. We follow mainly the review
\cite{review} using the notation and the sign convention accepted in  this paper. In $f(R)$ gravity, the action reads
%%%%%%
\be{2.1}
S=\frac{1}{2\kappa^2}\int d^4x\sqrt{-g}f(R) + S_m\, ,
\ee
%%%%%%
where $f(R)$ is an arbitrary smooth function of the scalar curvature $R$, $S_m$ is the action of matter, $\kappa^2=8\pi G_N$, and $G_N$ is the Newtonian
gravitational constant. The equation of motion corresponding to this action is (see, e.g., Eq. (2.4) in \cite{review})
%%%%%%
\ba{2.2} &{}&F(R)R_{\mu\nu} - \frac{1}{2}f(R)g_{\mu\nu} - \nabla_{\mu}\nabla_{\nu}F(R) + g_{\mu\nu}\square F(R)\nn\\
&=&{\kappa}^2 T_{\mu\nu}\, \quad \mu,\nu=0,1,2,3\, . \ea
%%%%%%
The trace of this equation gives
%%%%%%
\be{2.3}
3\square F(R)+F(R)R-2f(R)=\kappa^2 T\, .
\ee
%%%%%%
Here, $F(R)=f'(R)$ and $T=g^{\mu\nu}T_{\mu\nu}$. Besides, $\square F = (1/\sqrt{-g})\partial_{\mu}(\sqrt{-g}g^{\mu\nu}\partial_{\nu}F)$. In what follows,
%Greek indices $\mu,\nu=0,1,2,3$, the Latin indices $i,j=1,2,3$ and
the prime denotes the derivative with respect to the scalar curvature $R$. In our paper we consider a special class of $f(R)$ models which have solutions
$R_{\mathrm{dS}}$ of the equation
%%%%%
\be{2.4}
F(R)R-2f(R)=0\, .
\ee
%%%%%%%
As it follows from Eq. \rf{2.3}, they are vacuum solutions ($T=0$) of this equation for which the Ricci scalar is constant ($\square F(R)=0$). Such solutions
are called de Sitter points \cite{review,Od1,Od2}. According to \cite{osc1,osc3}, viable nonlinear models should have stable de Sitter points in the late
Universe. We can expand the function $f(R)$ in the vicinity of one of these points:
%%%%%
\ba{2.5} f(R)&=&f(R_{\mathrm{dS}})+f'(R_{\mathrm{dS}})(R-R_{\mathrm{dS}})+o(R-R_{\mathrm{dS}})\nn\\
&=&-f(R_{\mathrm{dS}})+ \frac{2f(R_{\mathrm{dS}})}{R_{\mathrm{dS}}}R + o(R-R_{\mathrm{dS}})\, , \ea
%%%%%%
where we used Eq. \rf{2.4}. Now, in order to have linear gravity at the late stage of the Universe evolution \cite{review}, without loss of generality we
choose the parameters of the model in such a way that
%%%%%
\be{2.6}
2\frac{f(R_{\mathrm{dS}})}{R_{\mathrm{dS}}}=1 \quad \Rightarrow \quad f(R_{\mathrm{dS}})=\frac{R_{\mathrm{dS}}}{2}\, .
\ee
%%%%%%
Therefore, we get
%%%%%
\be{2.7}
f(R)=-2\Lambda +R +o(R-R_{\mathrm{dS}})\, ,
\ee
%%%%%%
where $\Lambda \equiv R_{\mathrm{dS}}/4$. The stability of these points was discussed in \cite{review,Amendola}. Obviously, these models go asymptotically to
the de Sitter space when $R\to R_{\mathrm{dS}}\neq 0$ with a cosmological constant $\Lambda=R_{\mathrm{dS}}/4$. This happens when the matter content becomes
negligible with respect to $\Lambda$ as in the late Friedman-Robertson-Walker (FRW) cosmology. We can also consider a zero solution $R_{\mathrm{dS}}=0$ of Eq.
\rf{2.4}. It is more correct to call this point a Minkowski one. Here, $\Lambda =0$, and such models go asymptotically to the Minkowski space. In particular,
three popular models, Starobinsky \cite{Star2}, Hu-Sawicki \cite{HS} and MJWQ \cite{MJWQ}, have stable nonzero de Sitter points in future (approximately at the
redshift $z=-1$) \cite{JPS1,JPS2}. The explicit search for dS points in both future and past was considered in \cite{Od2,Od3}. It is worth noting that in
papers \cite{osc1,osc2} the authors point to the oscillating behavior of the parameter of the equation of state near the value $-1$ in the future. Moreover,
the number of times of such oscillations can be infinite \cite{osc3}.

In the case of the spatially flat background spacetime with the FRW metric
%%%%%
\be{2.8}
ds^2=g_{\mu\nu}dx^{\mu}dx^{\nu}=-dt^2+a^2(t) \left(dx^2+dy^2+dz^2\right)
\ee
%%%%%
and matter in the form of a perfect fluid with the energy-momentum tensor components $\bar T^{\mu}_{\nu}=\mathrm{diag}(-\bar\rho,\bar P,\bar P,\bar P)$, Eq.
\rf{2.2} results in the following system:
%%%%%
\be{2.9} 3FH^2 = (FR - f)/2 - 3H\dot{F} + \kappa^2 \bar \rho \, \ee
%%%%%%
and
%%%%%%
\be{2.10} -2F\dot{H}= \ddot {F} - H\dot{F} + \kappa^2 (\bar\rho + \bar P)\, , \ee
%%%%%
where the bar denotes the homogeneous background quantities, the Hubble parameter $H=\dot a/a$ (the dot everywhere denotes the derivative with respect to the
synchronous time $t$) and the scalar curvature
%%%%%
\be{2.11}
R=6\left(2H^2 + \dot{H}\right)\, .
\ee
%%%%%%
The perfect fluid  satisfies the continuity equation
%%%%%
\be{2.12}
\dot{\bar\rho} + 3H(\bar\rho + \bar P)=0\, ,
\ee
%%%%%%
which for nonrelativistic matter with $P=0$ has the solution
%%%%%%
\be{2.13}
\bar \rho = \bar\rho_c/a^3\, ,
\ee
%%%%%%
where $\bar\rho_c=\const$ is the rest mass density in the comoving  coordinates.
%\textcolor{red}{\bf Thus, we have briefly considered the homogeneous (flat for simplicity) background problem and established all necessary %designations.}

Above, Eqs. \rf{2.8}-\rf{2.13} describe the homogeneous background. As we have written in the Introduction, we consider the Universe at late stages of its
evolution when galaxies and clusters of galaxies have already formed and when the Universe is highly inhomogeneous inside the cell of uniformity which is
approximately 190 Mpc in size \cite{EZcosm2}. Obviously, these inhomogeneities perturb the homogeneous background. At scales larger than the cell of uniformity
size, the matter fields (e.g., cold dark matter) are well described by the hydrodynamical approach.  However, at smaller scales the mechanical approach looks
more adequate \cite{EZcosm2,EZcosm1}. In the framework of the mechanical approach galaxies, dwarf galaxies and clusters of galaxies (composed of baryonic and
dark matter) can be considered as separate compact objects. Moreover, at distances much greater than their characteristic sizes they can be described well as
point-like matter sources with the rest mass density
%%%%%%
\be{2.14}
\rho = \frac{1}{a^3}\sum_i m_i\delta{(\bf{r}-\bf{r}}_i)\equiv \rho_c/a^3\, ,
\ee
%%%%%%
where ${\bf{r}}_i$ is the radius-vector of the i-th gravitating mass in the comoving coordinates. This is the generalization of the well-known astrophysical
approach \cite{Landau} to the case of dynamical cosmological background. Usually, the gravitational fields of these inhomogeneities are weak and their peculiar
velocities are much less than the speed of light. All these inhomogeneities/fluctuations result in scalar perturbations of the FRW metric \rf{2.8}. In the
conformal Newtonian (longitudinal) gauge, such perturbed metric is \cite{review,Pt1,Pt2}
%%%%%%%
\be{2.15}
ds^2= -(1+2\Phi)dt^2+a^2(1-2\Psi)\left(dx^2+dy^2+dz^2\right)\, ,
\ee
%%%%%%%
where the introduced scalar perturbations $\Phi,\Psi \ll 1$. These functions of all spacetime coordinates, representing deviations of metric coefficients from
their average/background values, may be associated with famous Bardeen's potentials \cite{Pt1} under the made gauge choice. It is worth noting that smallness
of these nonrelativistic gravitational potentials $\Phi$ and $\Psi$ and smallness of peculiar velocities are two independent conditions (e.g., for very light
relativistic masses the gravitational potential can still remain small). Therefore, similar to the astrophysical approach described in \cite{Landau} (see \S
106), we split the investigation of galaxy dynamics in the late Universe into two steps. First, we neglect peculiar velocities and define the gravitational
potential $\Phi$. Then, we use this potential to determine dynamical behavior of galaxies. It gives us a possibility to take into account both the
gravitational attraction between inhomogeneities and the global cosmological expansion of the Universe. For example, for the linear model $f(R)=R$ this
procedure was performed in \cite{EKZ2}. Our present paper is devoted to the first step in this program. In other words, we are going to define scalar
perturbations $\Phi,\Psi$ for the $f(R)$ gravitational models. Under our assumptions and according to \cite{review,HN1,HN2}, these perturbations satisfy the
following system of equations:
%%%%%%
\ba{2.16}
&-&\frac{\Delta\Psi}{a^2} + 3H\left(H\Phi+\dot\Psi\right) = \nn \\
&-&\frac{1}{2F}\left[\left(3H^2+ 3 \dot{H} +\frac{\Delta}{a^2}\right)\delta F - 3H\dot{\delta F}\right. \nn\\
&+& \left.3H\dot{F}\Phi+3\dot{F} \left(H\Phi+\dot\Psi\right) +\kappa^2 \delta\rho\right]\, , \ea
%%%%%%%
\be{2.17}
H\Phi+\dot{\Psi} = \frac{1}{2F}\left(\dot{\delta F} - H \delta{F} - \dot{F}\Phi \right)\, ,
\ee
%%%%%%%%
\be{2.18}
-F(\Phi-\Psi)=\delta F\, ,
\ee
%%%%%%%%
\ba{2.19}
&{}& 3\left(\dot H\Phi+H\dot\Phi+\ddot\Psi\right) + 6H\left(H\Phi+\dot\Psi\right) + 3\dot H\Phi+\frac{\Delta\Phi}{a^2} \nn\\
&=&\frac{1}{2F}\left[3\ddot{\delta F}+3H\dot{\delta F}-6H^2\delta F - \frac{\Delta\delta F}{a^2} - 3\dot{F}\dot{\Phi}\right. \nn\\
&-& \left.3\dot{F}\left(H\Phi+\dot\Psi\right)- \left(3H\dot{F} + 6\ddot F\right)\Phi + \kappa^2\delta\rho\right]\, , \ea
%%%%%%%
\ba{2.20} &{}&\ddot{\delta F}+ 3H \dot{\delta F} - \frac{\Delta\delta F}{a^2}-\frac{1}{3}R\delta F\nn\\
&=&\frac{1}{3}\kappa^2(\delta \rho - 3\delta P) + \dot{F}(3H\Phi+3\dot\Psi+\dot{\Phi})\nn\\
&+&2\ddot F\Phi + 3H\dot{F}\Phi - \frac{1}{3}F\delta{R}\, , \ea
%%%%%%%
\ba{2.21} &{}&\delta F=F'\delta R,\quad \delta R=-2\left[3\left(\dot H\Phi+H\dot\Phi+\ddot\Psi\right)\right. \nn\\
&+&\left. 12H\left(H\Phi+\dot\Psi\right) +\frac{\Delta\Phi}{a^2}+3\dot{H}\Phi - 2\frac{\Delta\Psi}{a^2}\right]\, . \ea
%%%%%%%
In these equations the function $F$, its derivative $F'$ and the scalar curvature $R$ are unperturbed background quantities. Here $\Delta$ is the Laplacian in
the comoving coordinates. As a matter source, we consider dust-like matter. Therefore, $\delta P =0$ and
%%%%%
\be{2.22}
\delta \rho = \rho -\bar \rho =(\rho_c-\bar\rho_c)/a^3\, ,
\ee
%%%%%
where $\bar \rho$ and $\rho$ are defined in Eqs. \rf{2.13} and \rf{2.14}, respectively.

It can be easily verified that in the linear case $f(R)=R\ \Rightarrow\ F(R)=1$ this system of equations is reduced to Eqs. (2.18)-(2.20) in \cite{EZcosm1}.

%%%%%%%%%%%%%%%%%%%%%%%%%%%%%%%%%%%%%%%%%%%%%%%%%%%%%%%%%%%%%%%%%%%%%%%%%%%%%%%%%%%%%%%%%%%
%%%%%%%%%%%%%%%%%%%%%%%%%%%%%%%%%%%%%%%%%%%%%%%%%%%%%%%%%%%%%%%%%%%%%%%%%%%%%%%%%%%%%%%%%%%

\section{Astrophysical and cosmological approaches}

\setcounter{equation}{0}

\subsection{Astrophysical approach}

First, we consider Eqs. \rf{2.16}-\rf{2.21} in the astrophysical approach. This means that we neglect the time dependence of functions in these equations by
setting all time derivatives equal to zero. It is supposed also that the background model is matter free, i.e. $\bar\rho=0$. As we mentioned above, there are
two types of vacuum background solutions of Eq. \rf{2.3}: de Sitter spacetime with $R_{\mathrm{dS}}=12H^2=\const\neq 0$  and Minkowski spacetime with $R=0,\
H=0$. However, the system of equations \rf{2.16}-\rf{2.21} was obtained for the FRW metric \rf{2.15} where we explicitly took into account the dependence of
the scale factor $a$ on time. Therefore, if we want to get the time independent astrophysical equations directly from \rf{2.16}-\rf{2.21}, we should also
neglect the time dependence of $a$, i.e. the background Hubble parameter $H=0$. This means that the background solution is the Minkowski spacetime. This
background is perturbed by dust-like matter with the rest mass density \rf{2.14}. Keeping in mind that $\bar\rho=0$, we have $\delta \rho = \rho$.

\

In the case of Minkowski spacetime background, dropping the time derivatives, Eqs. \rf{2.16}-\rf{2.21} in the astrophysical approach are reduced to the
following system:
%%%%%%
\be{3.13}
-\frac{\Delta}{a^2}\Psi=-\frac{1}{2F}\left(\frac{\Delta}{a^2}\delta F + \kappa^2 \delta\rho\right)\, ,
\ee
%%%%%%%
%%%%%%
\be{3.14} -F(\Phi- \Psi)=\delta F\, ,
\ee
%%%%%%
%%%%%%
\be{3.15}
\frac{\Delta}{a^2}\Phi=\frac{1}{2F}\left(-\frac{\Delta}{a^2}\delta F + \kappa^2 \delta\rho\right)\, ,
\ee
%%%%%%
%%%%%%
\be{3.16} -\frac{\Delta }{a^2}\delta F=\frac{1}{3} \kappa^2\delta\rho -\frac{1}{3}F\delta R\, ,
\ee
%%%%%
%%%%%%
\be{3.17} \delta F=F'\delta R,\quad \delta R=-2\left(\frac{\Delta}{a^2}\Phi- 2\frac{\Delta}{a^2}\Psi\right)\, . \ee
%%%%%%

From \rf{3.13} and \rf{3.15} we obtain respectively
%%%%%%
\ba{3.18} \Psi&=&\frac{1}{2F}\delta F +\frac{\varphi}{a}=\frac{F'}{2F}\delta R +\frac{\varphi}{a}\, ,\nn\\
\Phi&=&-\frac{1}{2F}\delta F+\frac{\varphi}{a}=-\frac{F'}{2F}\delta R+\frac{\varphi}{a}\, , \ea
%%%%%%
where the function $\varphi$ satisfies the equation
%%%%%%
\be{3.19}
\Delta\varphi=\frac{1}{2F}\kappa^2a^3\delta\rho=\frac{1}{2F}\kappa^2\rho_c=\frac{4\pi G_N}{F}\rho_c\, .
\ee
%%%%%
Here, we took into consideration that in the astrophysical approach $\delta\rho_c=\rho_c$ where $\rho_c$ is defined by \rf{2.14}. It is worth noting that in
the Poisson equation \rf{3.19} the Newtonian gravitational constant $G_N$ is replaced by an effective one $G_{\mathrm{eff}}=G_N/F$.

Eq. \rf{3.14} follows directly from \rf{3.18} and therefore may be dropped, while from \rf{3.16} we get the following Helmholtz equation with respect to the
scalaron function $\delta R$:
%%%%%%
\be{3.20}
\Delta \delta R - \frac{a^2 F}{3 F'}\delta R = -\frac{a^2F}{3F'}\frac{\kappa^2}{Fa^3} \delta\rho_c\, .
\ee
%%%%%%%
On the other hand, it can be easily seen that the substitution of Eqs. \rf{3.18} and \rf{3.19} into \rf{3.17} results in the same Eq. \rf{3.20}. Therefore, in
the case of Minkowski background, the mass of the scalaron squared is
%%%%%
\be{3.21}
M^2 = \frac{a^2 }{3}\frac{F}{F'}\, .
\ee
%%%%%%
A similar formula (up to the evident substitution $a=1$ and the zero background scalar curvature) for the mass squared can be found, e.g., in
\cite{Starob,FL-T,Faraoni,EZf(R)} (see also Eq. (5.2) in the review \cite{review}).

\subsection{Cosmological approach}

Now we want to take into consideration cosmological evolution. This means that background functions may depend on time. In this case it is hardly possible to
solve the system \rf{2.16}-\rf{2.21} directly. Therefore, first, we study the case of the very large mass of the scalaron. It should be noted also that we
investigate the Universe filled with nonrelativistic matter with the rest mass density  $\bar \rho \sim 1/a^3$. Hence, we will drop all terms which decrease
(with increasing $a$) faster than  $1/a^3$. This is the accuracy of our approach. Within this approach, $\delta \rho \sim 1/a^3$ \cite{EZcosm1}.

\

{\em Large scalaron mass case}

\

As we can see from \rf{3.21}, the limit of the large scalaron mass corresponds to $F' \to 0$. Then, $\delta F$ is also negligible (see \rf{2.21}). Therefore,
Eqs. \rf{2.16}-\rf{2.21} read
%%%%%%
\ba{3.22} &{}&-\frac{\Delta\Psi}{a^2} + 3H\left(H\Phi+\dot\Psi\right)\nn\\
&=& -\frac{1}{2F}\left[ 3H\dot{F}\Phi+3\dot{F} \left(H\Phi+\dot\Psi\right) {+\kappa^2\delta\rho}\right]\, , \ea
%%%%%%
%%%%%%
\be{3.23} H\Phi+\dot{\Psi} = \frac{1}{2F}\left( - \dot{F}\Phi \right)\, ,\ee
%%%%%%
%%%%%%
\be{3.24} \Phi-\Psi=0\, , \ee
%%%%%%%
%%%%%%%
\ba{3.25}
&{}&3\left(\dot H\Phi+H\dot\Phi+\ddot\Psi\right) + 6H\left(H\Phi+\dot\Psi\right) + 3\dot H\Phi+\frac{\Delta\Phi}{a^2} \nn\\
&=&\frac{1}{2F}\left[- 3\dot{F}\dot{\Phi}-3\dot{F}\left(H\Phi+\dot\Psi\right)\right.\nn\\
&-& \left.\left(3H\dot{F} + 6\ddot F\right)\Phi {+\kappa^2\delta\rho}\right]\, , \ea
%%%%%%
%%%%%%%
\be{3.26} 0= \dot{F}(3H\Phi+3\dot\Psi+\dot{\Phi})+2\ddot F\Phi + 3H\dot{F}\Phi {+\frac{1}{3}\kappa^2\delta\rho-\frac{1}{3}F\delta R}\, ,\ee
%%%%%%%
%%%%%%%
\ba{3.27} &-&\frac{1}{2}\delta R=3\left(\dot H\Phi+H\dot\Phi+\ddot\Psi\right) + 12H\left(H\Phi+\dot\Psi\right)\nn\\
&+&\frac{\Delta\Phi}{a^2}+3\dot{H}\Phi - 2\frac{\Delta\Psi}{a^2}\, .\ea
%%%%%%%
From \rf{3.23} and \rf{3.24} we get
%%%%%%
\be{3.28} \Psi=\Phi=\frac{\varphi}{a\sqrt{F}}\, ,\ee
%%%%%%
where the introduced function $\varphi$ depends only on spatial coordinates. Substituting \rf{3.28} into \rf{3.22}, we obtain
%%%%%
\be{3.29} \frac{1}{a^3\sqrt{F}}\Delta \varphi+ \frac{{3\dot{F}}^2}{4aF^2\sqrt{F}}\varphi={\frac{1}{2F}\kappa^2\delta\rho}\, .\ee
%%%%%%%%

As we mentioned above, neglecting relativistic matter in the late Universe, we have $\delta\rho\sim1/a^3$ \cite{EZcosm1}. This approximation is getting better
and better performed in the limit $a\rightarrow+\infty$. We assume that this limit corresponds to the final stage of the Universe evolution. The similar limit
with respect to the scalar curvature is $R\rightarrow R_{\infty}$, where the value $R_{\infty}$ is just finite. Then, from \rf{3.29} we immediately come to the
condition
%%%%%
\be{3.30} F=\mathrm{const}+o(1)\, ,\ee
%%%%%%
where $o(1)$ is any decreasing (with increasing $a$) function of $a$. This condition holds at the considered late stage. One can also naively suppose that in
the late Universe $\dot F\sim1/a+o(1/a)$. However, as we will see below, this is wrong. Obviously, without loss of generality we can suppose that
$\mathrm{const}=1$. From the condition \rf{3.30} we get
%%%%%
\be{3.31} F=1+o(1) \quad \Rightarrow \quad f=-2\Lambda+R+o(R-R_{\infty})\, ,\ee
%%%%
where $\Lambda$ is the cosmological constant. Therefore, the limit of the large scalaron mass takes place for models which possess the asymptotic form
\rf{3.31}. For example, $R_{\infty}$ may correspond to the de Sitter point $R_{\mathrm{dS}}$ in future (see Eq. \rf{2.7}). As we have written in Sec. 2, all
three popular models, Starobinsky \cite{Star2}, Hu-Sawicki \cite{HS} and MJWQ \cite{MJWQ}, have such stable de Sitter points in future (approximately at the
redshift $z=-1$) \cite{JPS1,JPS2}. The condition of stability is $0<R F'/F< 1$ (see, e.g., (4.80) in \cite{review}). Since $F\approx 1$, this condition reads
$0<R< 1/F'$ which is fulfilled for the de Sitter points in the above-mentioned models. The reason of it consists in the smallness of $F'$.

We now return to the remaining Eqs. \rf{3.25}-\rf{3.27} to show that they are satisfied within the considered accuracy. First, we study \rf{3.25} which after
the substitution of \rf{3.28} and \rf{3.29} and some simple algebra takes the form
%%%%%
\be{3.32}
\frac{\varphi}{a}\dot H -\frac{\varphi}{2aF}\left(H\dot F- \ddot{F}\right) =0\, .
\ee
%%%%%%
To estimate $\dot F$ and $\ddot F$, we take into account that in the limit $R\to R_{\infty}$, $F\approx 1$, $H\approx \mathrm{const}\ \Rightarrow\ \dot H
\approx 0$, and $F'(R_{\infty})$ is some finite positive value. Then, $\dot F=F'\dot R\sim F'(R_{\infty})\dot R\sim \dot T \sim d\left(1/a^3\right)/dt\sim
H\left(1/a^3\right)\sim 1/a^3$ and $\ddot F \sim \dot a/a^4\sim 1/a^3$. Therefore, the left hand side of Eq. \rf{3.32} is of the order $o(1/a^3)$ and we can
put it zero within the accuracy of our approach. Similarly, Eqs. \rf{3.26} and \rf{3.27} are satisfied within the considered accuracy. It can be also seen that
the second term on the left hand side of Eq. \rf{3.29} is of the order $O(1/a^7)$ and should be eliminated.

Thus, in the case of the large enough scalaron mass we reproduce the linear cosmology from the nonlinear one, as it should be.

\

{\em Quasi-static approximation}

\

Now we do not want to assume a priori that the scalaron mass is large, i.e. $F'$ can have arbitrary values. Hence, we will preserve the $\delta F$ terms in
Eqs. \rf{2.16}-\rf{2.21}. Moreover, we should keep the time derivatives in these equations. Such a system is very complicated for the direct integration.
However, we can investigate it in the quasi-static approximation \cite{Ts,Ts2} (see also Sec. 8.1 in \cite{review}). According to this approximation, the
spatial derivatives give the main contribution to Eqs. \rf{2.16}-\rf{2.21}. Therefore, first, we should solve "astrophysical" Eqs. \rf{3.13}-\rf{3.17}, and
then check whether the found solutions satisfy (up to the adopted accuracy) the full system of equations. In other words, in the quasi-static approximation it
is naturally supposed that the gravitational potentials (the functions $\Phi,\,\Psi$) are produced mainly by the spatial distribution of
astrophysical/cosmological bodies. As we have seen, Eqs. \rf{3.13}-\rf{3.17} result in \rf{3.18}-\rf{3.21}. Now, we should keep in mind that we have the
cosmological background. Moreover, we consider the late Universe which is not far from the de Sitter point $R_{\mathrm{dS}}$ in future\footnote{Therefore, the
closer to $R_{\mathrm{dS}}$ we are, the more correct our approximation is.}. This means that $\delta \rho = \rho - \bar \rho$ in \rf{3.19}, all background
quantities (e.g., $F,\,F'$) are calculated roughly speaking at $R_{\mathrm{dS}}$ and the scalaron mass squared \rf{3.21} reads now
\cite{review,Od3,Starob,FL-T,Faraoni}
%%%%%
\be{3.33}
M^2 = \frac{a^2 }{3}\left(\frac{F}{F'}-R_{\mathrm{dS}}\right)\, .
\ee
%%%%%
Let us consider now Eq. \rf{3.20} with the mass squared \rf{3.33}. Taking into account that now $\delta\rho_c=\rho_c-\bar\rho_c$, we can rewrite this equation
as follows:
%%%%%%
\be{3.34}
\Delta \widetilde{\delta R} - M^2\widetilde{\delta R} + \frac{a^2}{3F'}\frac{\kappa^2}{a^3} \sum_i m_i\delta{(\bf{r}-\bf{r}}_i)=0\, ,
\ee
%%%%%%%
where
%%%%%
\be{3.35}
\widetilde{\delta R}= \delta R + \frac{a^2}{3F'M^2}\frac{\kappa^2}{a^3}\bar\rho_c=\delta R + \frac{\kappa^2}{(F-F'R_{\mathrm{dS}})a^3}\bar\rho_c\, .
\ee
%%%%%
Eq. \rf{3.34} demonstrates that we can apply the principle of superposition solving this Helmholtz equation for one gravitating mass $m_i$. Then the general
solution for a full system is the sum over all gravitating masses. As boundary conditions, we require for each gravitating mass the behavior $\widetilde{\delta
R}\sim 1/r$ at small distances $r$ and $\widetilde{\delta R}\to 0$ for $r\to\infty$. Taking all these remarks into consideration, we obtain for a full system
%%%%%%%
\ba{3.36} \delta R&=&\frac{\kappa^2}{12\pi aF'}\sum_i\frac{m_i\exp\left(-M {|\bf{r}-\bf{r}}_i|\right)}
{{|\bf{r}-\bf{r}}_i|}\nn\\
&-&\frac{\kappa^2}{(F-F'R_{\mathrm{dS}})a^3}\bar\rho_c \, . \ea
%%%%%%%
It is worth noting that averaging over the whole comoving spatial volume $V$ gives the zero value $\overline{\delta R}=0$. Really, since
$\sum_im_i/V=\bar\rho_c$,
%%%
\ba{m1} \overline{\delta R}&=&\frac{1}{V}\int\limits_V\delta RdV=\frac{1}{V}\frac{\kappa^2}{12\pi
aF'}\nn\\
&\times&\sum_im_i\frac{4\pi}{M^2}-\frac{\kappa^2}{(F-F'R_{\mathrm{dS}})a^3}\bar\rho_c=0\, .\ea
%%%
This result is reasonable because the rest mass density fluctuation $\delta\rho$, representing the source of the metric and scalar curvature fluctuations
$\Phi,\,\Psi$ and $\delta R$, has a zero average value $\overline{\delta\rho}=0$. Consequently, all enumerated quantities should also have zero average values:
$\bar\Phi=\bar\Psi=0$ and $\overline{\delta R}=0$, in agreement with \rf{m1}.

From Eq. \rf{3.18} we get the scalar perturbation functions $\Psi$ and $\Phi$ in the following form:
%%%%%%
\ba{3.37}
\Psi&=&\frac{F'}{2F}\left[\frac{\kappa^2}{12\pi aF'}\sum_i\frac{m_i\exp\left(-M {|\bf{r}-\bf{r}}_i|\right)}
{{|\bf{r}-\bf{r}}_i|}\right.\nn\\
&-&\left.\frac{\kappa^2}{(F-F'R_{\mathrm{dS}})a^3}\bar\rho_c\right]+\frac{\varphi}{a}\, ,\\
\label{3.38} \Phi&=&-\frac{F'}{2F}\left[\frac{\kappa^2}{12\pi aF'}\sum_i\frac{m_i\exp\left(-M {|\bf{r}-\bf{r}}_i|\right)}
{{|\bf{r}-\bf{r}}_i|}\right.\nn\\
&-&\left.\frac{\kappa^2}{(F-F'R_{\mathrm{dS}})a^3}\bar\rho_c\right]+\frac{\varphi}{a}\, , \ea
%%%%%%%
where $\varphi$ satisfies Eq. \rf{3.19} with $\delta\rho$ in the form \rf{2.22} (i.e. $\bar\rho_c\neq 0$). Obviously, when $F'\to 0$, $M\to \infty$, and we
have $\exp\left(-M {|\bf{r}-\bf{r}}_i|\right)/{|\bf{r}-\bf{r}}_i|\to 4\pi\delta({\bf r}-{\bf r}_i)/M^2$, so the expression in the square brackets in \rf{3.37}
and \rf{3.38} is equal to $\kappa^2\delta\rho_c/\left[(F-F'R_{\mathrm{dS}})a^3\right]$. Therefore, in the considered limit $F'\to 0$ we reproduce the scalar
perturbations $\Phi,\,\Psi$ from the previous large scalaron mass case, as it certainly should be.

Thus, neglecting for a moment the influence of the cosmological background, but without neglecting the scalaron's contribution, we have found the scalar
perturbations. They represent the mix of the standard potential $\varphi/a$ (see the linear case \cite{EZcosm1}) and the additional Yukawa term which follows
from the nonlinearity.

Now we should check that these solutions satisfy the full system \rf{2.16}-\rf{2.21}. To do it, we substitute \rf{3.36}, \rf{3.37} and \rf{3.38} into this
system of equations. Obviously, the spatial derivatives disappear. Keeping in mind this fact, the system \rf{2.16}-\rf{2.21} is reduced to the following
equations:
%%%%%%
\ba{3.39} &{}&3H\left(H\Phi+\dot\Psi\right) = -\frac{1}{2F}\left[\left(3H^2+ 3 \dot{H}\right)\delta F \right.\nn\\
&-& \left.3H\dot{\delta F} + 3H\dot{F}\Phi+3\dot{F} \left(H\Phi+\dot\Psi\right) \right]\, , \ea
%%%%%%
%%%%%%
\be{3.40} H\Phi+\dot{\Psi} = \frac{1}{2F}\left(\dot{\delta F} - H \delta{F} - \dot{F}\Phi \right)\, ,\ee
%%%%%%
%%%%%%
\ba{3.41} &{}& 3\left(\dot H\Phi+H\dot\Phi+\ddot\Psi\right) + 6H\left(H\Phi+\dot\Psi\right) + 3\dot H\Phi =\nn\\
&=&\frac{1}{2F}\left[3\ddot{\delta F}+3H\dot{\delta F}-6H^2\delta F  - 3\dot{F}\dot{\Phi}\right.\nn\\
&-& \left.3\dot{F}\left(H\Phi+\dot\Psi\right)- \left(3H\dot{F} + 6\ddot F\right)\Phi \right]\, , \ea
%%%%%%%
%%%%%%%
\be{3.42} \ddot{\delta F}+ 3H \dot{\delta F} = \dot{F}(3H\Phi+3\dot\Psi+\dot{\Phi})+2\ddot F\Phi + 3H\dot{F}\Phi \, ,\ee
%%%%%%%
%%%%%%%
\ba{3.43} &{}&\delta F=F'\delta R,\quad \frac{F'}{F}R_{\mathrm{dS}}\delta R=-2\left[3\left(\dot H\Phi+H\dot\Phi+\ddot\Psi\right) \right.\nn\\
&+&\left. 12H\left(H\Phi+\dot\Psi\right) +3\dot{H}\Phi \right]\, .\ea
%%%%%%%
Here, the term $-(R/3)\delta F$ (the term $(F'/F)R_{\mathrm{dS}}\delta R$) in the left hand side of \rf{3.42} (\rf{3.43}) disappears (appears) due to
redefinition of the scalaron mass squared \rf{3.33}.

All terms in \rf{3.36}, \rf{3.37} and \rf{3.38} depend on time, and therefore may contribute to Eqs. \rf{3.39}-\rf{3.43}. As we wrote above, according to our
nonrelativistic approach, we neglect the terms of the order $o(1/a^3)$. On the other hand, exponential functions decrease faster than any power function.
Moreover, we can write the exponential term in \rf{3.36} as follows:
%%%%%%
\be{3.44} \frac{\kappa^2}{12\pi F'}\sum_i\frac{m_i\exp\left(- \sqrt{\frac{1 }{3}\left(\frac{F}{F'}-R_{\mathrm{dS}}\right)}
|{\bf{r}}_{\mathrm{ph}}-{\bf{r}}_{\mathrm{ph}i}|\right)}{|{\bf{r}}_{\mathrm{ph}}-{\bf{r}}_{\mathrm{ph}i}|}\, , \ee
%%%%%%
where we introduced the physical distance $r_{\mathrm{ph}} = a r$. It is well known that astrophysical tests  impose  strong restrictions on the nonlinearity
\cite{NJ,BG} (the local gravity tests impose even stronger constraints \cite{EZf(R),NJ,BG}). According to these constraints, \rf{3.44} should be small at the
astrophysical scales. Consequently, on the cosmological scales it will be even much smaller. So, we will not take into account the exponential terms in the
above equations. However, in \rf{3.36}, \rf{3.37} and \rf{3.38}, we have also $1/a^3$ and $1/a$ terms which we should examine. Before performing this, it
should be recalled that we consider the late Universe which is rather close to the de Sitter point. Therefore, as we already noted in the previous subsection,
$F\approx 1$, $H\approx \mathrm{const}\ \Rightarrow\ \dot H \approx 0,\, R_{\mathrm{dS}}=12 H^2$ and $F'(R_{\mathrm{dS}})$ is some finite positive value.
Additionally, $\dot F,\ddot F,\dot F' \sim 1/a^3$. Hence, all terms of the form of $\dot F,\ddot F,\dot F' \; \times \; \Phi,\Psi,\dot \Phi, \dot\Psi$ are of
the order $o(1/a^3)$ and should be dropped. In other words, the functions $F$ and $F'$ can be considered as time independent.

First, let us consider the terms $\Psi =\Phi = \varphi/a$ in Eqs. \rf{3.37} and \rf{3.38} and substitute them  into Eqs. \rf{3.39}-\rf{3.43}. Such $1/a$ term
is absent in $\delta R$. So, we should put $\delta R=0$, $\delta F=0$. Obviously, this is the linear theory case. It can be easily seen that all equations are
satisfied. Indeed, the functions $\Phi$ and $\Psi$  are included in Eqs. \rf{3.39}-\rf{3.41} and \rf{3.43} (Eq. \rf{3.42} is satisfied identically) in
combinations $H\Phi+\dot\Psi$ and $H\dot \Phi+\ddot\Psi$ which are equal to zero.

Now, we study the terms $\sim 1/a^3$, i.e.
%%%%%
\ba{3.45} \delta R&=&-\frac{\kappa^2}{(F-F'R_{\mathrm{dS}})}\frac{\bar\rho_c}{a^3}\, ,\nn\\
\Psi&=&-\frac{\kappa^2F'}{2F(F-F'R_{\mathrm{dS}})}\frac{\bar\rho_c}{a^3}\, ,\nn\\
\Phi&=&\frac{\kappa^2F'}{2F(F-F'R_{\mathrm{dS}})}\frac{\bar\rho_c}{a^3}\, . \ea
%%%%%
Let us examine, for example, Eq. \rf{3.39}. Keeping in mind that $\delta F=F'\delta R$, one can easily get
%%%%%%
\ba{3.46} &{}&12H^2\frac{\kappa^2F'}{2F(F-F'R_{\mathrm{dS}})}\frac{\bar\rho_c}{a^3}\nn\\
&=& 12H^2\frac{\kappa^2F'}{2F(F-F'R_{\mathrm{dS}})}\frac{\bar\rho_c}{a^3}+o(1/a^3)\, . \ea
%%%%%%
Therefore, the terms $\sim 1/a^3$ exactly cancel each other, and this equation is satisfied up to the adopted accuracy $o(1/a^3)$. One can easily show that the
remaining Eqs. \rf{3.40}-\rf{3.43} are fulfilled with the same accuracy.

Thus, we have proved that the scalar perturbation functions $\Psi$ and $\Phi$ in the form \rf{3.37} and \rf{3.38} satisfy the system of Eqs.
\rf{2.16}-\rf{2.21} with the required accuracy. Both of these functions contain the nonlinearity function $F$ and the scale factor $a$. Therefore, both the
effects of nonlinearity and the dynamics of the cosmological background are taken into account. The function $\Phi$ corresponds to the gravitational potential
of the system of inhomogeneities. Hence, we can study the dynamical behavior of the inhomogeneities (e.g., galaxies and dwarf galaxies) including into
consideration their gravitational attraction and cosmological expansion, and also taking into account the effects of nonlinearity. For example, the
nonrelativistic Lagrange function for a test body of the mass $m$ in the gravitational field described by the metric \rf{2.15} has the form (see \cite{EZcosm1}
for details):
%%%%%
\be{3.47}
L \approx   -m\Phi + \frac{m a^2{\bf{v}}^2}{2}\, ,\quad {\bf{v}}^2=\dot x^2 + \dot y^2 + \dot z^2\, .
\ee
%%%%%
We can use this Lagrange function for analytical and numerical study of mutual motion of galaxies. In the case of the linear theory, such investigation was
performed, e.g., in \cite{EKZ2}. With the help of the explicit expression \rf{3.38} we can perform now similar numerical and analytical investigations for
different $f(R)$ models.

%%%%%%%%%%%%%%%%%%%%%%%%%%%%%%%%%%%%%%%%%%%%%%%%%%%%%%%%%%%%%%%%%%%%%%%%%%%%%%
%%%%%%%%%%%%%%%%%%%%%%%%%%%%%%%%%%%%%%%%%%%%%%%%%%%%%%%%%%%%%%%%%%%%%%%%%%%%%%%%

\section{Conclusion}

In our paper we have studied scalar perturbations of the metric in nonlinear $f(R)$ gravity. The Universe has been considered at the late stage of its
evolution and at scales much less than the cell of uniformity size which is approximately 190 Mpc \cite{EZcosm2}. At such distances, our Universe is highly
inhomogeneous, and the averaged hydrodynamic approach does not work. Here, the mechanical approach \cite{EZcosm2,EZcosm1} is more adequate. Therefore, we have
used the mechanical approach to investigate the scalar perturbations in nonlinear theories. We have considered a class of viable $f(R)$ models which have de
Sitter points in future with respect to the present moment \cite{osc1,osc3,osc2}.

The main objective of this paper was to find explicit expressions for $\Phi$ and $\Psi$ in the framework of nonlinear  $f(R)$ models. Unfortunately, in the
case of nonlinearity the system of equations for scalar perturbations is very complicated. It is hardly possible to solve it directly. Therefore, we have
considered the following approximations: the astrophysical approach, the large scalaron mass case and the quasi-static approximation. In all three cases we
found the explicit expressions for the scalar perturbation functions $\Phi$ and $\Psi$ up to the required accuracy. The latter means that, because we consider
nonrelativistic matter with the averaged rest mass density $\bar\rho \sim 1/a^3$, all quantities in the cosmological approximation are also calculated up to
the corresponding orders of $1/a$. It should be also noted that in the cosmological approach our consideration is valid for nonlinear models where functions
$f(R)$ have the stable de Sitter points $R_{\mathrm{dS}}$ in future with respect to the present time, and the closer to $R_{\mathrm{dS}}$ we are, the more
correct our approximation is. All three popular models, Starobinsky \cite{Star2}, Hu-Sawicki \cite{HS} and MJWQ \cite{MJWQ} (see also \cite{Od2,Od3}) have such
stable de Sitter points in future (approximately at the redshift $z=-1$) \cite{JPS1,JPS2}.

The quasi-static approximation is of most interest from the point of view of the large-scale structure investigations. Here the gravitational potential $\Phi$
\rf{3.38} contains both the nonlinearity function $F$ and the scale factor $a$. Hence, we can study the dynamical behavior of the inhomogeneities (e.g.,
galaxies and dwarf galaxies) including into consideration their gravitational attraction and the cosmological expansion, and also taking into account the
effects of nonlinearity. All this makes it possible to carry out the numerical and analytical analysis of the large-scale structure dynamics in the late
Universe for nonlinear $f(R)$ models.

%This is the main result of our paper.

%%%%%%%%%%%%%%%%%%%%%%%%%%%%%%%%%%%%%%%%%%%%%%%%%%%%%%%%%%%%%%%%%%%%%%%%%%%%%%%%
%%%%%%%%%%%%%%%%%%%%%%%%%%%%%%%%%%%%%%%%%%%%%%%%%%%%%%%%%%%%%%%%%%%%%%%%%%%%%%%%

\section*{Acknowledgements}

The work of M. Eingorn was supported partially by NSF CREST award HRD-1345219 and NASA grant NNX09AV07A. M. Eingorn also thanks the DAAD for the scholarship
during the research visit to Cologne and the hospitality of the University of Cologne during the final preparation of this paper.

%%%%%%%%%%%%%%%%%%%%%%%%%%%%%%%%%%%%%%%%%%%%%%%%%%%%%%%%%%%%%%%%%%%%%%%%%%%%%%%%%%%%%%%
%%%%%%%%%%%%%%%%%%%%%%%%%%%%%%%%%%%%%%%%%%%%%%%%%%%%%%%%%%%%%%%%%%%%%%%%%%%%%%%%%%%%%%%

\end{document}